# Domain wall motion in a polycrystalline vortex lattice

Malcolm Durkin[1,*], Emily Waite[1], Taylor L. Hughes[1], Nadya Mason[2]
[1]*Department of Physics and Materials Research Laboratory, University of Illinois, Urbana, Illinois 61801, USA*
[*]*Current address: Department of Physics, University of Colorado Boulder, Boulder, Colorado 80305, USA*
[2]*Pritzker School of Molecular Engineering, University of Chicago, Chicago, Illinois 60637, USA*

Disorder fundamentally reshapes how crystalline systems respond to external forces, yet it remains unclear whether disorder drives interacting lattices toward glassy states or instead fragments them into domains separated by mobile interfaces. Here, we investigate vortex motion in superconducting island arrays, where disorder is introduced in a controlled manner by tuning the magnetic field away from commensurate vortex fillings. By driving vortices with an applied current, we observe a two-step depinning transition at incommensurate fillings. Comparison with molecular vortex model simulations shows that this intermediate regime is consistent with domain wall motion in a polycrystalline vortex lattice. While two-step depinning has been explored theoretically in driven periodic systems, direct experimental evidence linking this behavior to interface-dominated vortex motion has been lacking. Our results demonstrate that disordered, interacting vortex systems with strong periodic pinning can favor interface physics over homogeneous glassy dynamics.

## I. INTRODUCTION

Understanding how ordered structures lose their integrity in the presence of disorder is a central problem across the physical sciences. Crystals rarely fail all at once. Instead, disorder can fracture an ordered state into domains separated by interfaces, or it can drive the system into an amorphous, glass-like phase. Which of these outcomes emerges, and how the system responds when driven, strongly influences transport, dissipation, and mechanical response in systems ranging from atomic solids and magnetic materials to colloids and electronic matter [1–3].

A particularly rich setting in which to study these questions is provided by driven periodic systems, where interacting particles move through a landscape of repeating energy minima. In such systems, the competition between particle-particle interactions, the underlying periodic potential, and externally applied forces can give rise to complex nonequilibrium behavior. Depending on the degree and nature of disorder, motion may occur collectively, through localized defects, or via sliding interfaces that separate ordered regions [4–8]. Distinguishing between these regimes is essential for understanding how dissipation emerges in systems that are nominally ordered.

Magnetic vortices in type-II superconductors offer a powerful and tunable realization of this physics. In an ideal clean superconductor, vortices form an ordered Abrikosov lattice [9,10], but imperfections, interactions, and external driving forces can destabilize this arrangement. Vortex motion is not merely a microscopic curiosity: it directly produces electrical resistance and plays a key role in phenomena such as superconductor–insulator transitions, anomalous metallic states, and decoherence in superconducting quantum circuits [11–14]. Despite decades of study, however, it remains an open question whether disorder in interacting vortex systems preferentially produces glassy states or instead fragments the lattice into multiple crystalline domains separated by mobile interfaces [3,15–20].

One reason this question has been difficult to answer experimentally is the challenge of controllably introducing disorder. In bulk superconductors, disorder is often fixed by material growth and difficult to tune independently of other parameters. Two-dimensional superconducting systems with engineered periodic potentials offer a way forward. In these systems, disorder can be introduced geometrically, by tuning the mismatch between the natural spacing of vortices and the underlying potential landscape. This class of systems is closely related to the Frenkel–Kontorova model, which predicts a rich variety of static and dynamic phases depending on commensurability and driving force [4–6].

In this work, we exploit this strategy using superconductor–normal–superconductor (SNS) island arrays, which impose a well-defined periodic potential on vortices while allowing their density to be precisely controlled by an external magnetic field [21,22]. When the number of vortices is commensurate with the array geometry, vortices form highly ordered configurations that are strongly pinned, producing characteristic features in transport measurements such as resistance minima and enhanced depinning currents [22–27]. When the filling is shifted away from these special values, additional vortices act as defects, frustrating global crystalline order. By driving vortices with an applied current and measuring the resulting electrical response, we directly probe how disorder alters the onset and nature of vortex motion.

Our central experimental observation is that incommensurate vortex configurations exhibit a two-step depinning process. As the drive increases, vortices first enter an intermediate dynamical regime characterized by partial motion, before transitioning to full lattice flow at higher currents. This behavior contrasts sharply with commensurate configurations, which de-pin in a single step, as previously observed in ordered vortex lattices [6,16,25,26]. By comparing our transport measurements with molecular-dynamics simulations of interacting vortices in a periodic potential [24,28–30], we show that this intermediate regime is consistent with domain wall motion in a polycrystalline vortex lattice, rather than with glassy dynamics.

These results demonstrate that, in the presence of a sufficiently strong periodic potential, disorder in an interacting vortex system can favor interface-dominated transport, where motion is concentrated along the boundaries between ordered domains, as predicted in theoretical and numerical studies of driven periodic systems [30,31]. While two-step vortex depinning has been discussed theoretically, direct experimental evidence linking this behavior to domain wall motion in a polycrystalline vortex lattice has been lacking. More broadly, our findings clarify how crystalline order breaks down under frustration in driven periodic systems and provide experimental access to regimes that have previously been explored primarily through theory and simulation. This insight is directly relevant to understanding dissipation in superconducting circuits [11,12], nonequilibrium phases in two-dimensional superconductors [14,32], and emergent vortex behavior in engineered quantum materials such as Moiré systems [33,34].

## II. METHODS

Samples were fabricated on Si substrates with 500-1000 nm $SiO_2$ on the surface. Standard photolithography procedures were used to create a 4-point device geometry pattern, in which 1 nm Ti (for adhesion) and 10 nm Au were deposited via electron beam deposition. Note the Ti does not contribute to the superconducting state since overlaying normal metal (Au in this case) causes reverse proximity effect, suppressing the superconducting state in Ti [42]. The island pattern was created using standard electron beam lithography techniques, and 70 nm of Nb was deposited using a custom electron beam evaporation chamber.

## III. RESULTS

The devices we measure consist of triangular arrays of superconducting Nb islands on normal metal films in a four-point measurement configuration. For details on device fabrication, see the Methods section below. The triangular array consists of Nb islands that are 70 nm thick, 260 nm in diameter, and have 490 nm edge-to-edge spacing (i.e. a lattice parameter of 750 nm). The islands begin superconducting ~ 8 K, and were previously found to have a dirty-limit coherence length of $\xi_0^{\mathrm{Nb}} \approx 27$ nm [35]. Transport measurements show that the entire array becomes superconducting below a transition temperature $T_c$ [35], which depends on island spacing, with a zero-field transition that can be associated with a Berezinski-Kosterless-Thouless transition [29,36]. The array $T_c$ is shown in Fig. 1 (a). Measurements in this paper are performed at temperatures well below $T_c$, where the array exhibits magnetoresistance oscillations at finite fields consistent with vortex transport [29]. Previous studies of this superconducting island system reported that the critical current of individual islands is estimated to be ~100 μA at 30 mK, while the global critical current of the entire array was directly measured to be around 20 μA [29]. Our measurements did not exceed 9 μA of current put through the system, keeping the system well within the superconducting regime.

Samples are studied using four-point measurements (Fig. 1 (a) inset) in a dilution refrigerator at 17 mK, sweeping DC current and measuring voltage. Vortices experience a periodic potential from the island array [22], where the local energy minimum is at the center of each triangle formed between adjacent islands, and an energy barrier exists at the array edges. By current biasing the devices, the current applies a Lorentz force on the vortices. If sufficient to overcome the energy barrier, the Lorenz force will de-pin vortices and drive their motion, which is measured as a voltage across the sample. The number of vortices per triangle is determined by the magnetic field and island spacing, and is characterized by the number of flux quantum per plaquette (the "filling fraction"), or $f = \Phi/\Phi_0$, where $\Phi$ is the flux through a unit cell and $\Phi_0 = h/2e$ is the quantum of flux. The maximum magnetic field used was 8.5 mT, which corresponds to a filling factor of $f = 0.525$. The de-pinning and dynamic motion of these vortices in island arrays has previously been imaged using $\Phi_0$-magnetic force microscopy [20].

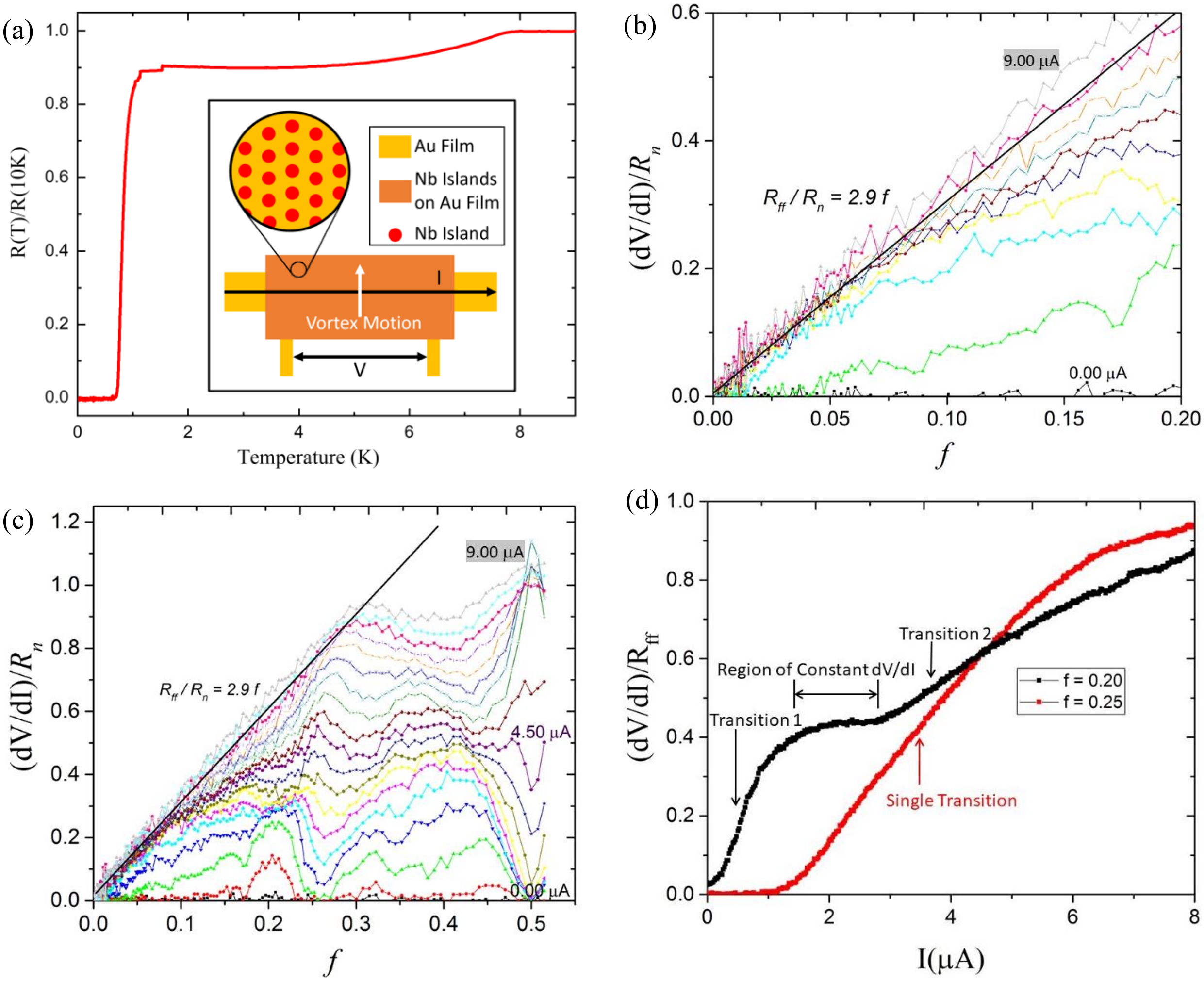

FIG. 1. *I-V* measurements. (a) Resistance vs temperature for a 490 nm edge-to-edge spaced array, showing the superconducting transition. Inset shows the four-point measurement configuration. (b) Normalized differential resistance $dV/dI$ at 1 $\mu A$ current intervals as a function magnetic filling *f*. As current increases at low fillings, there is a rapid transition from $dV/dI = 0$ to flux flow resistance (black line, $R_{ff}/R_N$). (c) $dV/dI$ vs *f* at 0.5 μA current intervals, for a larger range than (b). At higher fillings, dips associated with commensurate fillings are visible at $f = 1/6, 1/4, 1/3,$ and $1/2$. Intermediate clusterings of lines are visible at $f = 0.20$ and $f = 0.35$. (d) $dV/dI$ vs *I*. The black curve for $f = 0.20$ (incommensurate filling) undergoes a two-step transition with an intermediate region of constant $dV/dI$ in between. The red curve for $f = 0.25$ (commensurate filling) undergoes a single transition.

The dilute filling behavior of the arrays (i.e., at small $f$ values) is shown in Fig. 1 (b), a plot of $dV/dI$ as a function of $f$ for different $I$ (obtained by taking the derivative of an $I$-$V$ measurement). In this regime, the array undergoes a current driven transition from pinned vortices ($dV/dI = 0$) to flux flow (constant $dV/dI$) at a depinning current [6,25]. The flux flow—or lattice flow—regime occurs when all vortices are moving at a terminal velocity, where the Lorentz force is equal to a dissipative force, resulting in a linear relationship between $I$ and $V$. This leads to a constant $dV/dI$ that we refer to as the flux flow resistance, $R_{ff}$, which reflects uniform vortex motion across the array. $R_{ff}$ scales linearly with filling fraction $f$, as the measured $V$ is proportional to the total number of moving vortices. This is shown in Fig. 1 (b), where, for sufficient currents and low $f$ values, the $dV/dI$ curves fit to a single black line representing $R_{ff}$. When plotted on a broader range in Fig. 1 (c), the extrapolated fit to $R_{ff}$ represents an upper limit to the $dV/dI$ measured for a given $f$ and indicates when all vortices are flowing.

At larger $f$ values, $dV/dI$ no longer has a linear relationship with $f$ for most applied currents. This is visible in Fig. 1 (b), when a number of curves start to diverge from the $R_{ff}$ fit at $f > 0.05$; the effect is even more pronounced in Fig. 1 (c), where a pattern of peaks and dips emerge when $f > 0.1$. The departure from linear $dV/dI$ occurs because greater vortex density results in stronger vortex-vortex interactions. Vortex-vortex repulsion usually leads to weaker pinning as filling is increased. However, at special fillings—e.g., $f = 1/12, 1/6, 1/4, 1/2$—the vortex lattice is commensurate with the array potential wells, resulting in crystalline vortex orderings and strong pinning, as evidenced by dips in $dV/dI$ and a greater depinning current at commensurate fillings.

Commensurately and incommensurately filled lattices exhibit different dynamic behavior. The dynamics as a function of current can be seen in Fig. 1 (d), which plots $dV/dI$ vs $I$ for the commensurate filling $f = 0.25$ alongside the incommensurate pinning $f = 0.20$. Driven from pinned to flux flow, the commensurate filling undergoes a single transition, while the incommensurate filling undergoes two distinct transitions separated by an intermediate region of constant $dV/dI$. Similar behavior can be seen in Fig. 1 (b), where incommensurate fillings have intermediate clusterings of lines, as exemplified at $f = 0.20$. In contrast the commensurate fillings transition rapidly into flux flow, resulting in peak reversals where low current dips turn into peaks at higher currents (e.g., as at $f = 0.5$). Similar peak reversals have been observed in previous works [16] and used as evidence for a vortex Mott-insulator to metal transition [6,25,26,37–39].

To characterize incommensurate lattice dynamics, we plot the transition locations as a function of magnetic field and current in Fig. 2. Since the transitions are associated with steps in $dV/dI$, they can be identified as $d^2V/dI^2$ peaks, or the bright regions of Fig. 2. For commensurate fillings, depinning transitions occur at higher currents, as indicated by the arrows in Fig. 2. While dilute fillings and commensurate fillings only have a single transition (only one visible $d^2V/dI^2$ peak for a given $f$), incommensurate fillings often undergo a two-step transition with two visible $d^2V/dI^2$ peaks. This is evident in Fig. 2, where the incommensurate fillings indicated by dashed lines have first transitions marked by blue circles and secondary transitions marked by red X's. The peaks and dips in Fig. 2 are independent of magnetic field history, as the vortices find an energy minima in each magnetic field [40] (see Supplement for additional measurements on the history dependence [41]).

The single transitions smoothly split into two transitions as filling is shifted away from commensurate values. This can be seen in Fig. 2, where the single transition at $f = 1/6$ smoothly splits into two diverging transition curves as the filling is increased, with the upper curve marked with an "X" and the lower curve marked with an "O" at $f \sim 0.21$. Another prominent splitting is also visible as the field is increased from $f = 1/4$ and a less prominent secondary transition curve splits off as the filling is decreased from $f = 1/6$. The smooth splitting as a function of filling indicates that this behavior is determined by lattice structure, with the vortex lattice transitioning from pinned regime to an intermediate vortex motion regime to a lattice flow regime. This splitting has not been previously discussed and implies that vortex motion begins with interstitial vortices superimposed on the commensurate array, as explained below.

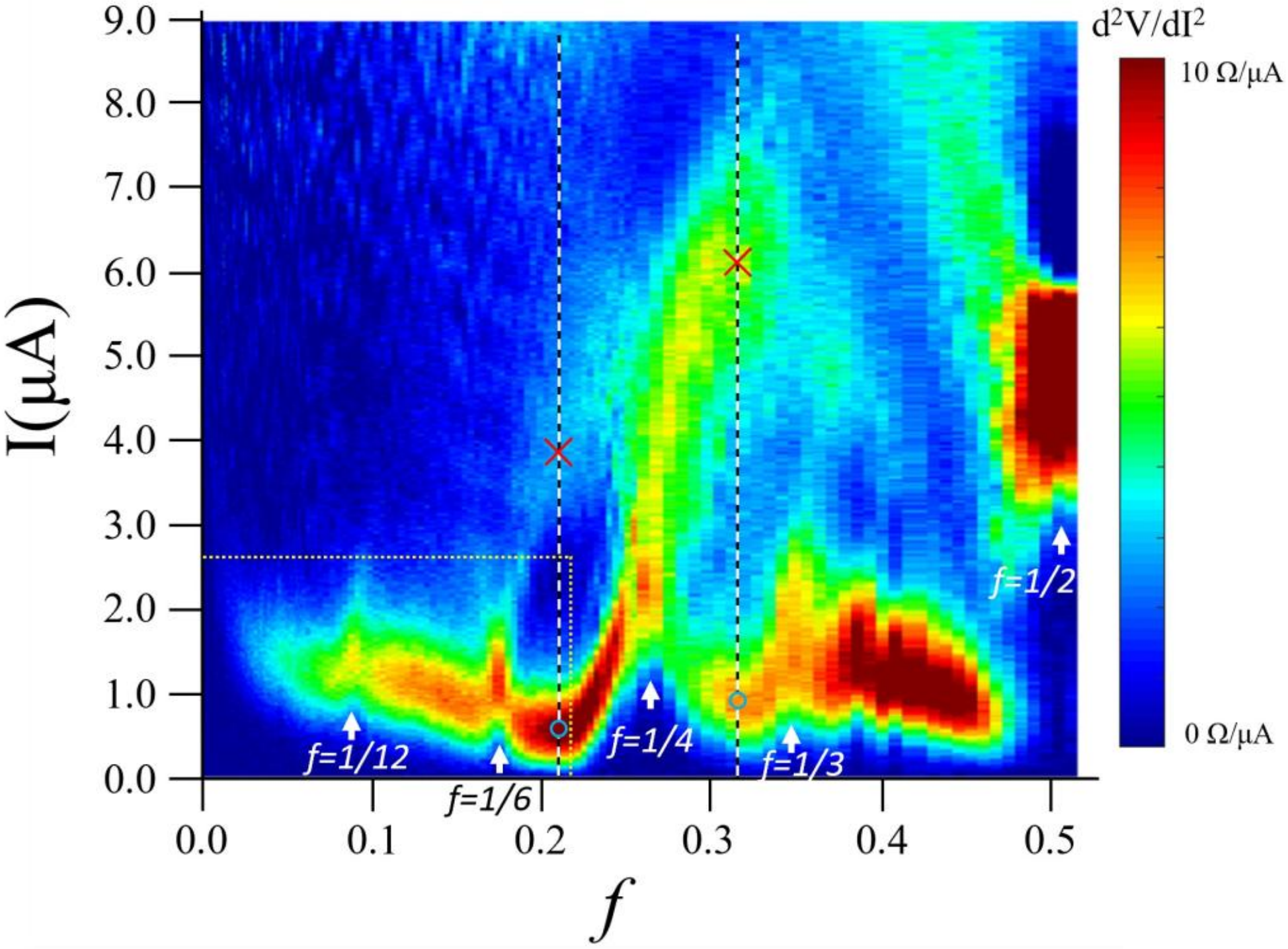

FIG. 2. $d^2V/dI^2$ as a function of current and frustration**.** Transition steps can be mapped using $d^{2V}/dI^2$, with higher values (brighter colors) corresponding to a transition step. Commensurate fillings are marked with white arrows and undergo a single transition. Examples of incommensurate fillings are marked with a dashed line and undergo two transitions, the first marked with a blue circle and the second marked with a red X. The region marked in a yellow dashed box in the lower left corner is simulated in Figure 4 (a).

## IV. DISCUSSION

To better understand the dynamic behavior in the different regimes, including the $d^2V/dI^2$ peak splitting, we use a molecular vortex model based on the Langevin equation for mutually repulsive vortices in a periodic potential [24]. This has equations of motion given by

$$m\ddot{x}_i = F_{applied} - \frac{\partial V\big(x_i(t)\big)}{\partial x_i} - \eta\dot{x}_i(t) + \varepsilon_i + \sum_{j=1}^{N} U\left(\frac{x_i(t) - x_j(t)}{L_{int}}\right), \quad (1)$$

where $m$ is a mass term related to capacitance, $F_{applied}$ is the Lorentz force (proportional to the applied current), $V\big(x_i(t)\big)$ is a periodic potential defined by the array, $\varepsilon_i(t)$ is a stochastic force used to simulate finite temperature, and $U(x)$ is the mutual repulsion between vortices. Low resistance systems such as this one are in the overdamped limit, with $m = 0$ [25]. Further details are found in the supplement [41]. By varying the number of vortices per potential well and applying a driving force, we simulate current sweeps at magnetic field intervals. This model has been previously used to describe the behavior of SNS arrays. Although standard vortex models predict a $dV/dI$ peak at the depinning current, we previously showed that this peak can be suppressed—while producing a non-zero linear *I–V* intercept—by including a history-dependent dissipative term in the simulations [29]. This hysteretic dissipative force arises from the time it takes for the vortex lattice to relax in the presence of moving vortices, thus causing vortices to be correlated in time [29].

A basic one-dimensional (1D) simulation, effectively the 1D FK model, can replicate the two-step transition for incommensurate fillings. Shown in Fig. 3 (a), the lattice undergoes a direct transition from pinned to flux flow at commensurate fillings, such as $f = 0.5$. Due to the ordered arrangement of vortices (Fig. 3 (a) upper inset), the vortex lattice moves in unison and exhibits dynamics similar to that of a single vortex. As additional vortices are added, they enter as “defects” of the commensurate vortex lattice (Fig. 3 (a) lower inset), separating ordered domains. These defects require a lower depinning force than the rest of the lattice and begin moving prior to the lattice depinning current, resulting in a transition from pinned to defect flow to lattice flow. For a more direct comparison with data, the predicted $dV/dI$ is provided in Fig. 3 (b) for the different transport regimes. Notably, the distinct two-step transition occurs when well-defined defects separate ordered regions, analogous to domain walls. In contrast, higher values of disorder than shown in Fig. 3 yield an amorphous state, exhibiting only a single transition. For a detailed analysis, see the Supplement [41].

To simulate a 2D system, we use a periodic potential similar to the one produced by the triangular island array, incrementally sweep current at different vortex populations, and extract the vortex motion. As shown in Fig. 4 (a), this simulation reproduces the basic features of the data in Fig. 2: depinning current peaks at commensurate fillings and a second $d^2V/dI^2$ peak that splits as the filling is altered from commensurate values. Similar to our measurements, the two-step transition shows clearly above $f = 1/6$, with the first transition marked with a navy-blue circle and a second transition marked by a grey X. There are some features missed by the simulation, such as the peak at $f = 1/12$, meaning that the model is incomplete (this is likely due to the existence of longer-range interactions between vortices not completely captured in the model). Furthermore, this model could not be applied to frustration larger than $f = 1/4$ due to model limitations and computational resources required; this may be addressed in future work.

The lattice structure is clarified by Fig. 4 (b), which shows simulated vortex positions and dynamics in the domain wall motion regime (marked by a white circle in Fig. 4 (a)). The black circles and red circles show initial and final vortex locations, respectively, over the short time roughly corresponding to a vortex crossing from one well to another (for a plot of lattice flow, which is above the second transition, see supplement [41]). Vortex motion occurs primarily in defects, which appear as cracks that form between crystalline structures. Motion along these cracks is visible as the line following the dark circles in Fig. 4 (b). The presence of defect lines not only indicates that the intermediate transport regime involves defect motion but is consistent with a polycrystalline structure where defects appear on the interfaces separating crystalline domains. These results suggest the intermediate transport regime involves domain wall motion.

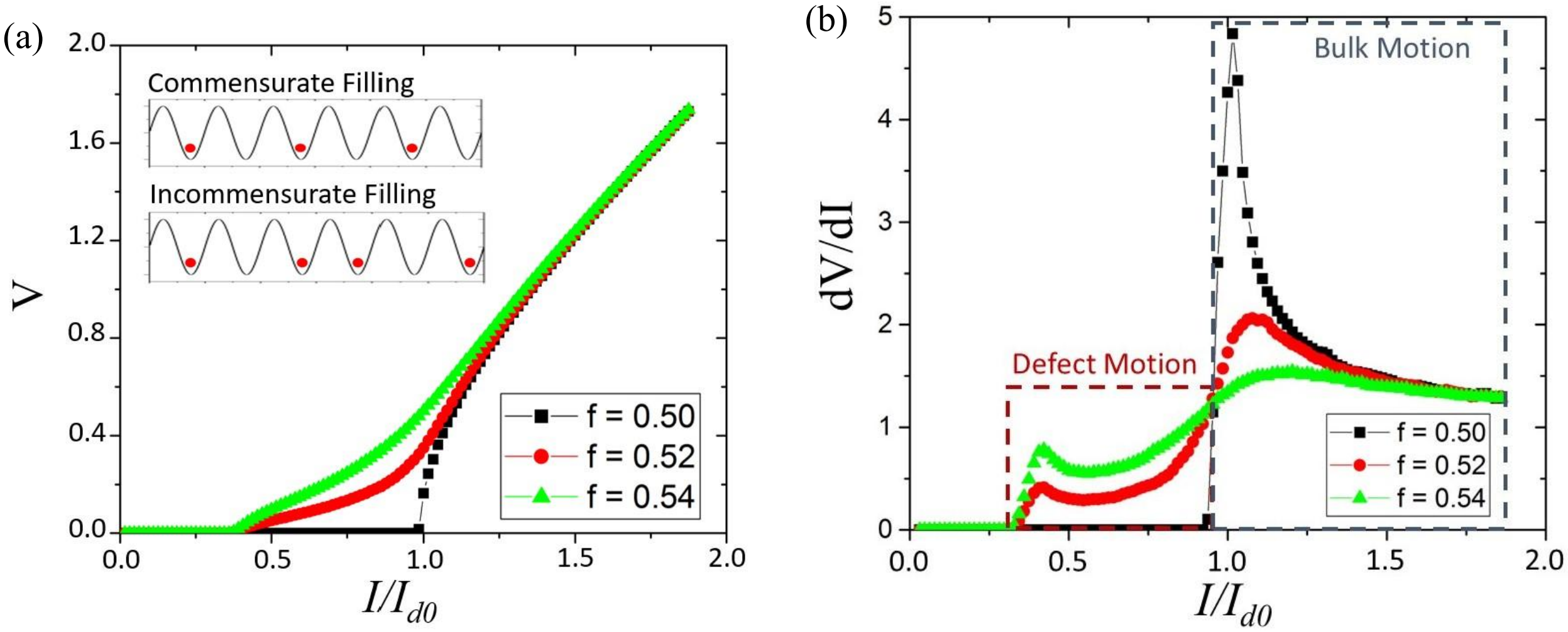

FIG. 3. One-dimensional simulation of array. (a) Simulated *I-V* behavior for commensurate ($f = 0.5$) and incommensurate fillings ($f = 0.52, 0.54$). Top inset shows the half filling arrangement. Bottom inset shows an incommensurate filling with defects in the form of vortices in adjacent wells. (b) Simulated $dV/dI$ measurements. At commensurate values, the system rapidly transitions from pinned to bulk vortex motion. As field is increased ($f = 0.52, 0.54$), defects are added. This results in a transition from pinned to defect motion to bulk vortex motion. $I_{d0}$ is the de-pinning current in the single particle limit.

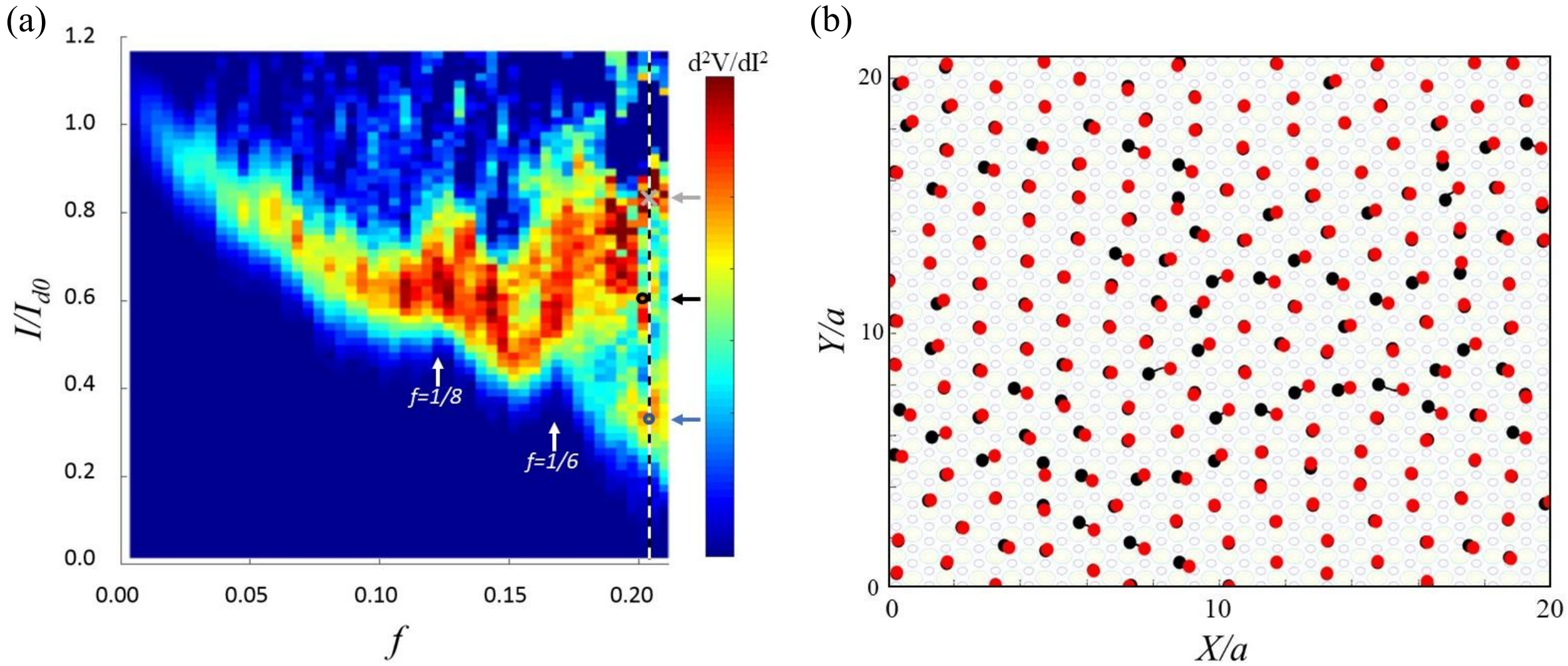

FIG. 4. 2D simulation of triangular array. (a) Simulated $d^2V/dI^2$ within region indicated by yellow dashed box in Figure 2. At $f = 1/6$, a second transition splits off from the de-pinning current when filling is increased. An incommensurate filling is shown with a dashed line. A blue circle at $I/I_{d0} \approx 0.3$ shows the first transition. A grey X at $I/I_{d0} \approx 0.8$ shows the second transition. $I_{d0}$ is the de-pinning current in the single particle limit. Arrows on the right indicate the position of the grey X and blue circle. (b) Time evolution of vortex motion at $f = 0.2$ with an applied field perpendicular to the vortex plane shown and an applied current of $I = 0.6\ I_{d0}$ , for data shown as a black circle in (a). Black solid circles are simulated vortices and red circles show their position a short time afterwards (with a black line showing the path in between). This simulation is in the defect motion regime, with most defect motion occurring on the interface between 2 different crystalline structures.

Thus, the two-step transition we observe at incommensurate fillings is consistent with a transition from pinned vortices to lattice defect motion to lattice flow. Molecular vortex model simulations suggest that this motion occurs on the edge of crystalline domains, providing evidence for domain wall motion and polycrystalline structure in this system similar to those recently predicted in artificial pinning array structures [27,30,31], but not previously explored in experimental studies.

## V. CONCLUSION

In this work, we investigated vortex dynamics in superconducting island arrays, focusing on the transition between commensurate and incommensurate vortex configurations. By carefully tuning the frustration between commensurate fractional values and incommensurate irrational values, and comparing the results to simulations of vortex systems, we found that our data is consistent with the formation of a polycrystalline vortex lattice, where vortex motion primarily occurs along domain walls. This study provides new insight into vortex lattice melting in systems with periodic pinning potentials. A deeper understanding of these vortex dynamics has implications for dissipation mechanisms in superconducting quantum circuits, anomalous metallic states in two-dimensional superconductors, and vortex behavior in Moiré materials.

## ACKNOWLEDGEMENTS

The fabrication of this project was carried out in part in the Material Research Laboratory

Central Research Facilities, University of Illinois. This work was primarily supported by the DOE "Quantum Sensing and Quantum Materials" Energy Frontier Research Center under Grant No. DE-SC0021238 and under National Science Foundation Grant No. NSF-2422090. Samples were fabricated with support from DOE Basic Energy Sciences under Grant No. DE-SC0012649. Transport measurements were performed with support from the U.S. Department of Energy, Office of Science, National Quantum Information Science Research Centers under Grant No. 9J-60012-0005A. The authors would like to acknowledge assistance in the transport measurements from Alexander Beach.

## Supplemental Material for "Domain wall motion in a polycrystalline vortex lattice"

### S1. Simulating Vortex-Vortex Repulsion

The magnitude of the vortex-vortex repulsion force is given by $U(X)=K_1(X/L_{int})$, where $K_1(x)$ a modified Bessel function of the second kind. In the long interaction limit ($L_{interaction}>>a$, where $a$ is center to center island spacing), which is applicable for arrays, this approaches $U(X)=C/X$ where $C$ is a constant modifying the magnitude of the repulsion. In order to have the calculations scale in a reasonably efficient way, interactions between vortices over $10a$ away are ignored. To avoid artifacts from vortices entering and leaving interaction ranges of other vortices, we smoothly tapper the interaction range using the following form:

$$U(x) = BU_0(x) = B\left(\frac{a}{x\left[1+exp\left(\frac{|x|-8a}{2a}\right)\right]} - \frac{1}{10(1+e)}\right) \tag{S1}$$

As shown in figure S1, this is a reasonable approximation for nearest neighbor and next nearest neighbor interactions in the regime of interest, while effectively ignoring interactions further than that.

### S2. One-Dimensional Simulation

The one-dimensional system is given by the potential,

$$V(x) = \frac{Aa}{2\pi}\sin\left(\frac{2\pi x}{a}\right) - F_I x \tag{S2}$$

Where $A$ controls barrier height and $F_I$ is the force resulting from the applied current. The system being modeled has low resistance and is in the overdamped limit, allowing $m$ to be set to $0$. The differential equation for an overdamped vortex array is then solved using Euler's method. This is an iterative method that repeats the following calculation over short periods of time, Δt,

$$\dot{x}_i(t=\Delta t) = \frac{1}{\eta}\left(F_I - A\cos\left(\frac{2\pi x_i(t)}{a}\right) + o_\iota(t) + B\sum_{J=1}^{N} U_0\left(x_\iota(t) - x_J(t)\right)\right). \tag{S3}$$

$$x_i(t+\Delta t) = x_i(t) + \dot{x}_\iota(t)\Delta\text{t}. \tag{S4}$$

The free parameter in this equation is the ratio between the periodic potential parameter, $A$, and the repulsion parameter $B$, a ratio that determines the stiffness of the lattice. The stochastic force, $\grave{o}_\iota(t)$, is obtained using a random number generator with an exponential distribution of the form

$$\exp\left(-\frac{|\epsilon_i(t)|}{kT}\right). \tag{S5}$$

The simulation seeds $N$ vortices in 50 wells ($f = N/50$) and then performs a slow anneal from high temperature to low temperature to find a ground state configuration, imposing periodic boundary conditions. We then run the simulation starting with this ground state at varying currents. The results can be seen in Figure S2 with two lattice stiffness parameters: $B/A = 2$ and $B/A = 6$. A stiffness of $B/A = 2$ results in broad two step regions around $f = 0.5$ and $f = 1.0$ (dark and light blue). In contrast, a stiffness of $B/A = 6$ has a distinct two step transition only in relatively narrow range around $f = 0.5$ and $f = 1.0$.

The relationship between vortex lattice structure and the presence (or absence) of a two-step transition is investigated in Figure S3, which shows the vortex structure at both $B/A = 2$ and $B/A = 6$. At the $f$ values given, $B/A = 2$ exhibits a prominent two step transition and $B/A = 6$ either exhibits a less visible two step transition or has only a single step, providing a comparison between the two observed transport phenomena. $B/A = 2$ [Figure S3 (a)(c)] results in well-defined defects that are limited in area to one or two wells: $f = 0.54$ yields defects in the form of vortices in adjacent wells and $f = 0.9$ yields defects in the form of empty wells in an otherwise filled lattice. In both cases, the defects can be interpreted as domain walls separating ordered domains, with $f = 0.54$ defects each separating two half-filled domains and the $f = 0.9$ defects each separating two entirely filled lattices.

In contrast, $B/A = 6$ defects are more difficult to identify spatially. Rather than appear as a pair of adjacent vortices or as an empty well, these defects are groupings of perturbed vortices that no longer rest in the center of the wells. As seen in Figure S3 (b)(d), the defects occur over a region 10 wells wide, with vortices either perturbed towards the center of the defect ($f = 0.54$) or away from the center of the defect ($f = 0.90$). At filling $f = 0.54$, there are some segments of vortices that are unperturbed, allowing for a visible intermediate step. At $f = 0.9$, the defects are close enough for the lattice to take on a quasiperiodic structure and there is only a single transition. Thus, the two step transition is a signature of distinct defects, which form on the interfaces between ordered domains.

**S3. Absence of a differential resistance peak in experimental data**

While a differential resistance peak is predicted in the simulations, it is absent in experimental simulations. We have previously addressed the associated I-V behavior using a history dependent dissipative force. To understand the nature of the peak, we included a history dependent dissipative force for a 1D system with $B/A = 2$, as shown in figure S4. Here, the inclusion of the term removes the peaks on both the defect motion and the lattice motion steps (1). Since this does not fundamentally alter the types of vortex motion occurring, we do not include this term in any other section of this work for simplicity and to save computational resources.

**S4. Two-Dimensional Simulation**

We create a potential with triangular barriers as well as exclusion zones where the superconducting islands would be using the form

$$V_1(x,y) = \exp\left(-\frac{\left(\mathrm{mod}\left(\frac{x}{a}-\frac{1}{4},1\right)-\frac{1}{2}\right)^2+\left(\mathrm{mod}\left(\frac{y}{a}+\frac{\sqrt{3}}{4},\sqrt{3}\right)-\frac{\sqrt{3}}{2}\right)^2}{2\sigma^2}\right) + \exp\left(-\frac{\left(\mathrm{mod}\left(\frac{x}{a}+\frac{1}{4},1\right)-\frac{1}{2}\right)^2+\left(\mathrm{mod}\left(\frac{y}{a}-\frac{\sqrt{3}}{4},\sqrt{3}\right)-\frac{\sqrt{3}}{2}\right)^2}{2\sigma^2}\right) \quad \text{(S6)}$$

$$V_2(x,y) = -\left|\sin\left(\frac{4\pi y}{a\sqrt{3}}\right)+\sin\left(\frac{2\pi}{a}\left(x+\frac{y}{\sqrt{3}}\right)\right)+\sin\left(\frac{2\pi}{a}\left(x-\frac{y}{\sqrt{3}}\right)\right)\right|^{N_{\mathrm{shape}}} \quad \text{(S7)}$$

$$V(x) = A\big(C_1V_1(x,y)+C_2V_2(x,y)\big) - F_I x \quad \text{(S8)}$$

where $V_1(x,y)$ is an exclusion zone for islands and $V_2(x,y)$ is the vortex barrier in between islands, with $C_1$ and $C_2$ setting the relative strengths of the two potentials. The $N_{\mathrm{shape}}$ parameter sets the shape of the vortex barrier as shown in figure S5(a), with $N_{\mathrm{shape}} = 1$ resulting in a broad potential well. We instead use the parameter $N_{\mathrm{shape}} = 4$ to get a narrower well. Setting $C_1 = 150$, $C_2 = 2/250$, and $\sigma = 0.2$, we get the potential shown in figure S5 (b) with the path along which the vortices move shown in white. The vortices follow a path from the center of an island triangle and cross the barrier through the center of the edge of a triangle, demonstrating that they follow the intended path.

Euler's method is once again used, in the overdamped limit, only broken up into $x$ and $y$ components.

$$\dot{x}_i(t) = \frac{1}{\eta}\left(-\frac{\partial V(x_i(t), y_i(t))}{\partial x_i} + \grave{o}_{xi}(t) + B\sum_{j=1}^{N} \frac{\left(x_i(t) - x_j(t)\right)}{r_{ij}(t)} U_0\left(r_{ij}(t)\right)\right) \quad \text{(S9)}$$

$$\dot{y}_i(t) = \frac{1}{\eta}\left(-\frac{\partial V(x_i(t), y_i(t))}{\partial y_i} + \grave{o}_{yi}(t) + B\sum_{j=1}^{N} \frac{\left(y_i(t) - y_j(t)\right)}{r_{ij}(t)} U_0\left(r_{ij}(t)\right)\right) \quad \text{(S10)}$$

It is then solved by performing the following operations repeatedly with periodic boundary conditions

$$x_i(t + \Delta t) = x_i(t) + \dot{x}_i(t)\Delta t, \quad \text{(S11)}$$

$$y_i(t + \Delta t) = y_i(t) + \dot{y}_i(t)\Delta t, \quad \text{(S12)}$$

where $r_{ij}$ is the distance between vortices $i$ and $j$. N vortices are randomly seeded into 240 wells and then slowly annealed into a low energy configuration. The results when $B/A$ = 8 are shown in figure 4 in the main text, with the intermediate motion type also shown. The lattice flow motion at $f$ = 0.20 are shown in figure S5. Fig. S5 and Fig. 4(b) were both taken 10,000 time steps after the current was set, to allow structure changes due to the driving current to occur. The time between the two snapshots is approximately the time necessary for a vortex to cross from one pinning site to another, given by $\Delta t = 1.5 t_I$ in Fig. 4(b) and $\Delta t = 1.0\ t_I$ in Fig. S6, where $t_I = 2\eta/3F_I$.

**S5. Reproducibility and history-dependence of measurements**

In order to ensure that the observed effects are reproducible and that the vortex dynamics do no change substantially, we have performed additional measurements on a similar sample (Same niobium on gold, however this sample had a 550 nm lattice constant). To test this, we first compared zero field cooled with field cooled data, with an applied field of 0.5 T, well above the critical field of the array. The results are shown in Fig. S7. We observe that close to the commensurate $f = 1/4$ frustration, the difference between field cooled and zero field cooled results are minimal. Away from the commensurate frustration, (for example, the $f = 0.22$ curve in Fig. S7), we see a shift in the shape of the differential conductance curve. This is likely due to imperfections in the vortex lattice causing some vortices to become very strongly pinned during field cooldown, which will then disrupt the beginning of the depinning transition in the vortex lattice. This feature is more pronounced in the incommensurate frustration since at the first flux flow transition only sees a small fraction of the vortices de-pinning, so any strongly pinned vortices will more strongly affect the transport measurement observed.

We observe a similar effect when testing the effects of ramping the field up versus down. After zero field cooling, we started at $f = 0.20$ and ramped up to $f = 0.25$. Then, we turned the field up to $f = 0.30$ and approached $f = 0.25$ from above. The results of this are shown in Fig. S8a-c. We can see that we get the same general features of a single depinning transition occurring, however the current at which this transition occurs shifts. This is likely due to small imperfections in the vortex lattice. While approaching from above, there may be additional remnant vortices that cause the frustration to be closer to exactly ¼, compared to when we approach from below. Additionally, we repeated the same procedure but went to a slightly incommensurate frustration ($f = 0.27$) and observe no discernable shift in the differential resistance.

**Supplementary Materials References**

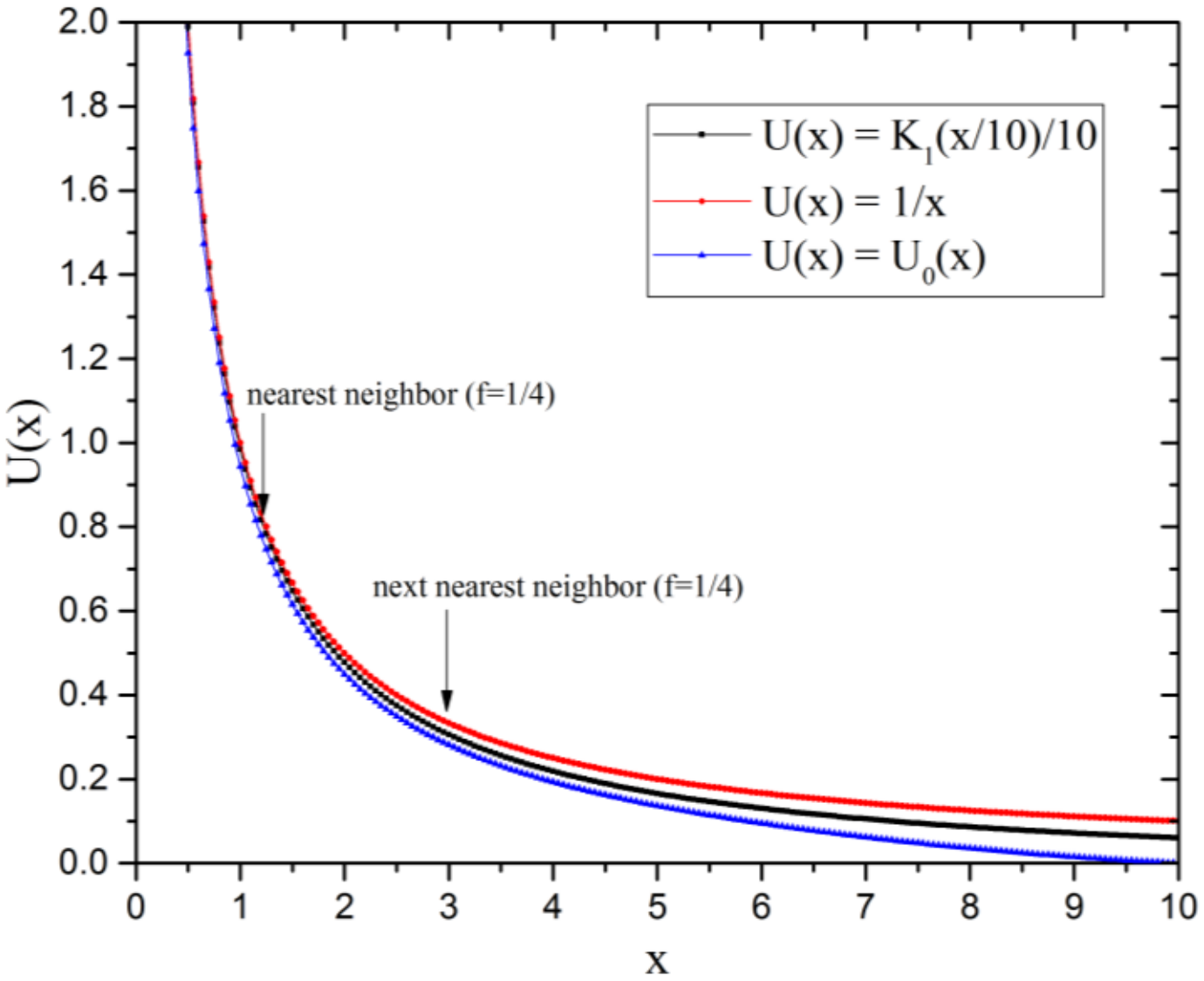

**Figure S1. Vortex-Vortex Repulsion as a function of distance** The different functions discussed as repulsion terms are shown. The function $U_0(x)$ given by equation S1 is a decent approximation of the predicted repulsion terms for nearest neighbor and next nearest neighbor interactions.

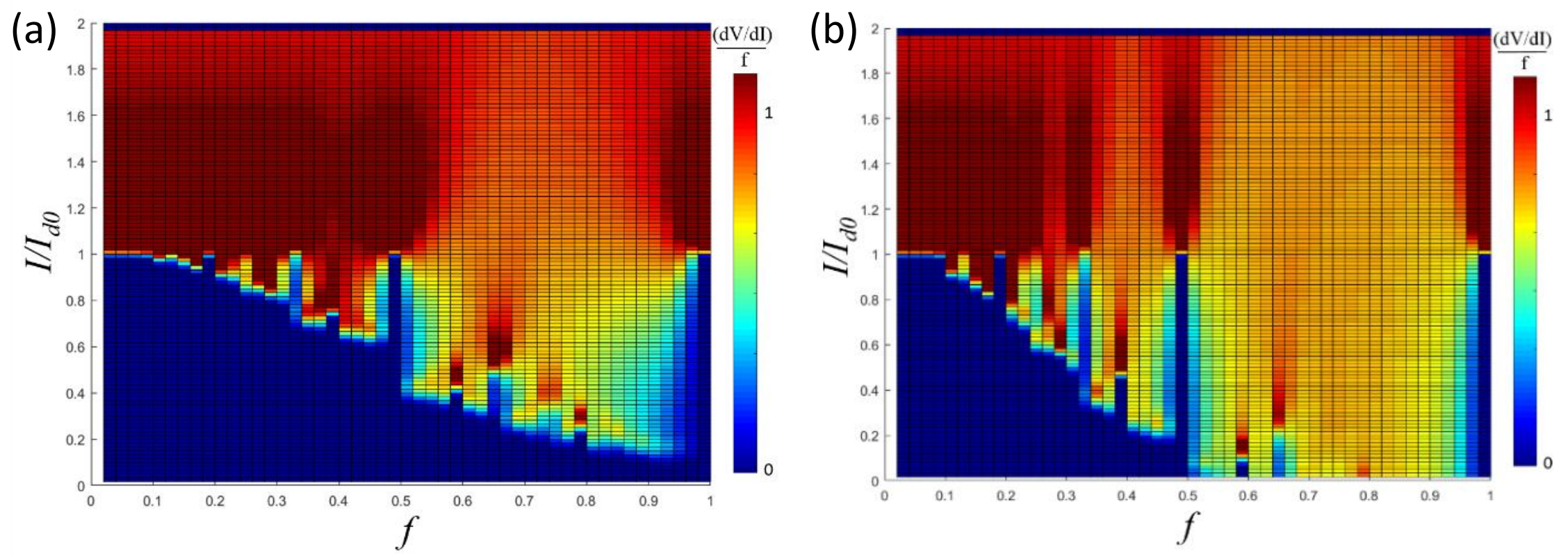

**Figure S2. Simulated 1D dV/dI as a function of current and field**. **(a)** $B/A = 2.0$, which is used in Figure 3, has visible intermediate steps associated with defect motion near commensurate fillings, most prominently just above $f = 0.5$ and just below $f = 1.0$. **(b)** $B/A = 6.0$ has a much stiffer lattice, resulting in narrower regions with two steps. Instead, the stiffer lattice favors lattice motion.

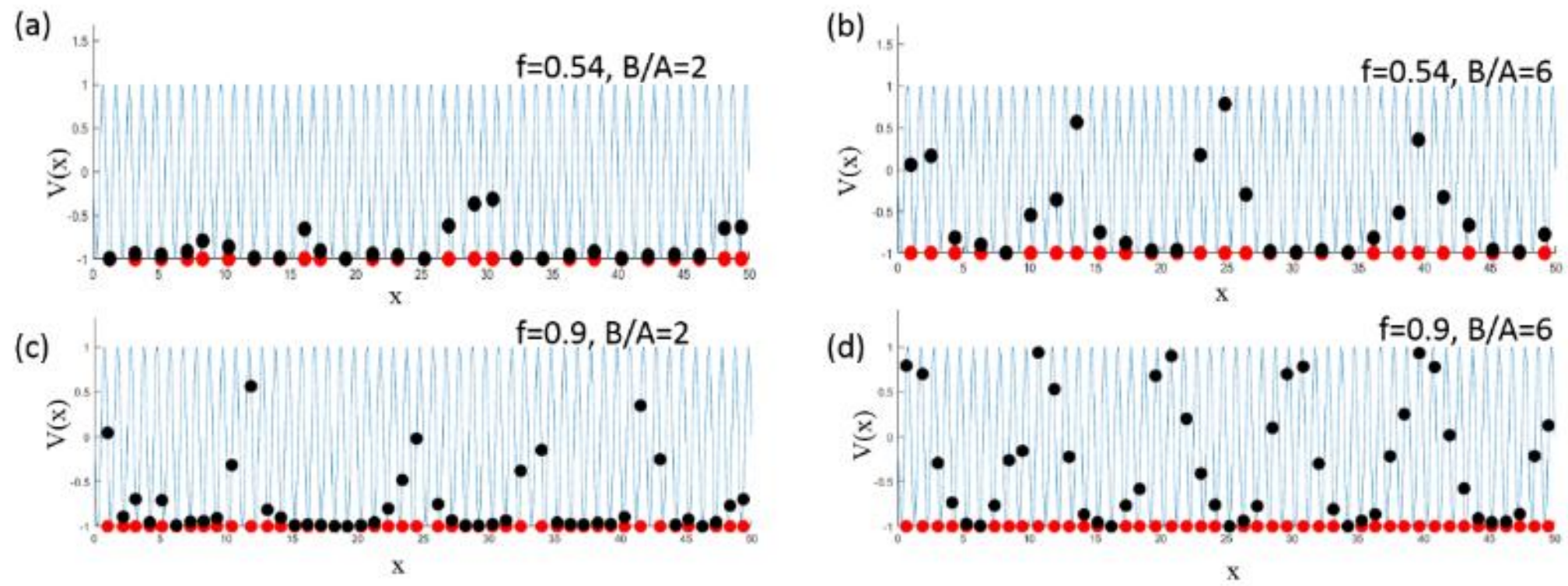

**Figure S3. Simulated 1D vortex arrangement as a function of stiffness and repulsion.** Blue lines represent the periodic potential, the red dots at the bottom of each graph show the x position of vortices, and the black dots show the energy of the vortices as a function of x position. **(a)** $f = 0.54$ and $B/A = 2.0$ yields defects in the form of a pair of vortices in adjacent wells. These defects separate ordered regions with half the wells occupied. **(b)** $f = 0.54$ and $B/A = 6.0$ vortices do not sit in the wells and are not separated by integer well periods. Disorder appears to manifest in a quasiperiodic structure rather than in the interface between two ordered domains as in (a). **(c)** has defects appear as empty wells separating domains with every well filled. **(d)** Greater vortex-vortex repulsion once again yields a quasiperiodic structure. It is notable that (a) and (c) exhibit a two-step transition in figure S2 (a), but (b) and (d) undergo a single transition in figure S2(b).

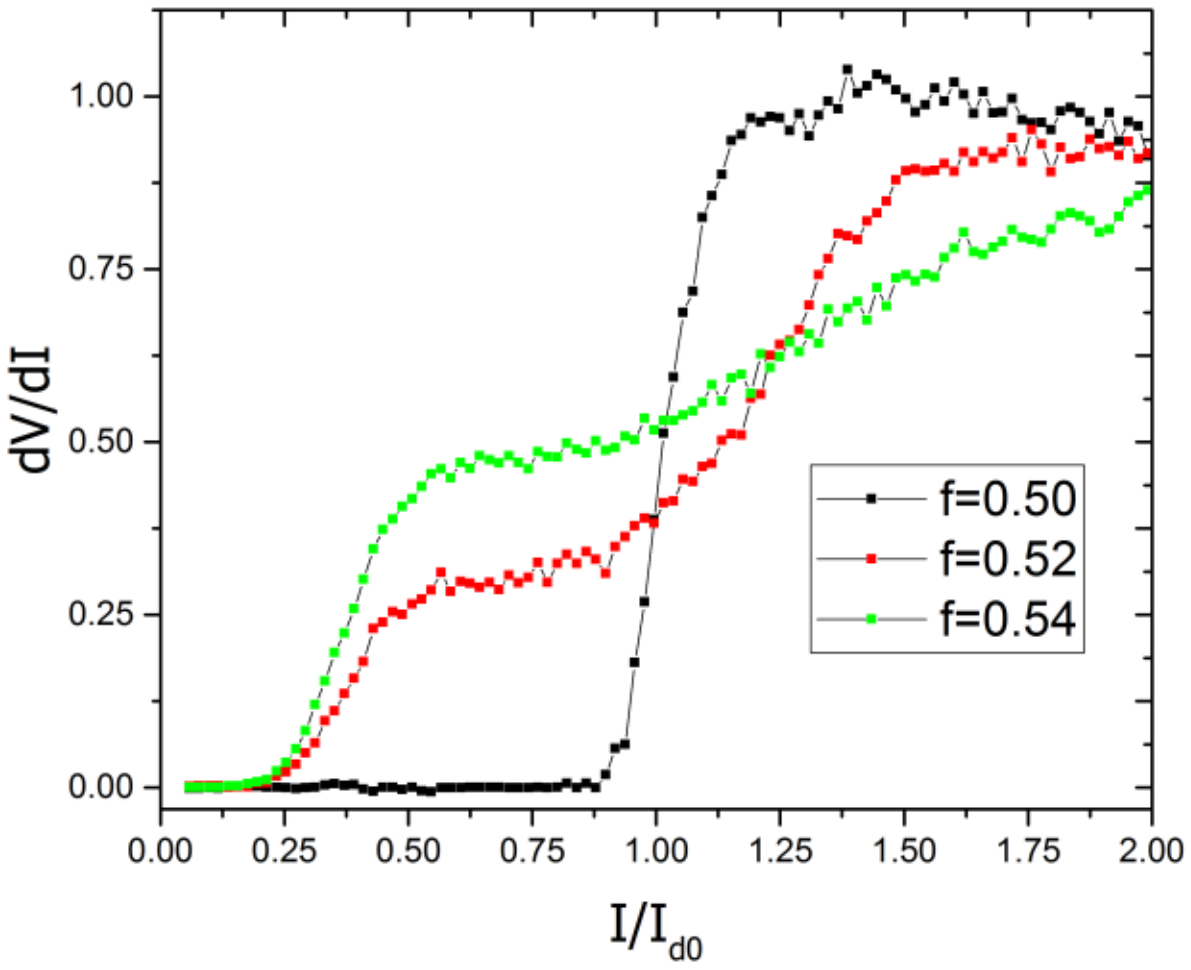

**Figure S4. Simulated 1D dV/dI with history dependent dissipative force.** The array undergoes a single transition from pinned to lattice flow for commensurate fillings. It undergoes a two-step transition from pinned to defect motion to lattice flow for incommensurate fillings. The inclusion of a history dependent dissipative force removes the differential resistance peak at each step.

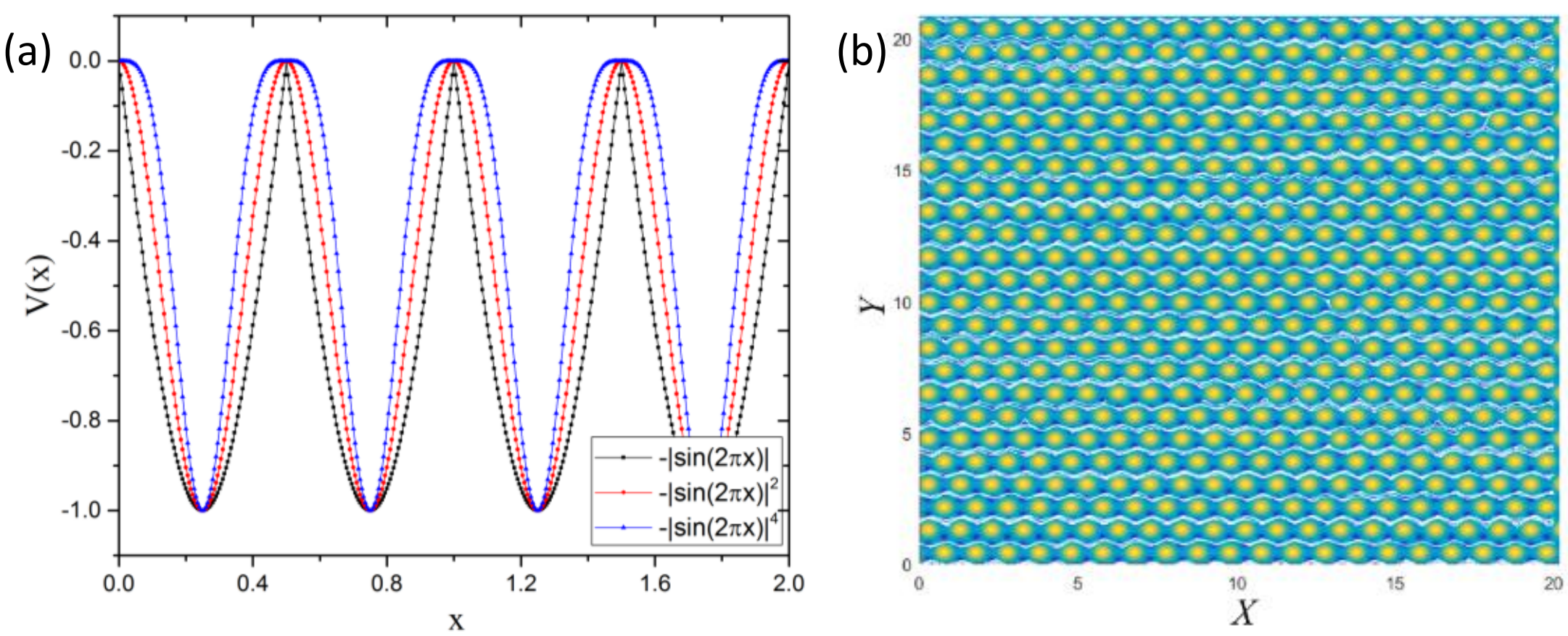

**Figure S5. 2D Potential Well. (a)** The effect of $N_{shape}$ on periodic potential. Higher values of $N_{shape}$ yield narrower wells for vortices to rest in. **(b)** The resulting periodic potential is shown with wells in blue and islands in yellow. The path of vortices in a simulated lattice is shown in white. The exclusion potential around the islands results in vortices moving between the centers of the adjacent wells, which requires overcoming the potential provided by equation S7.

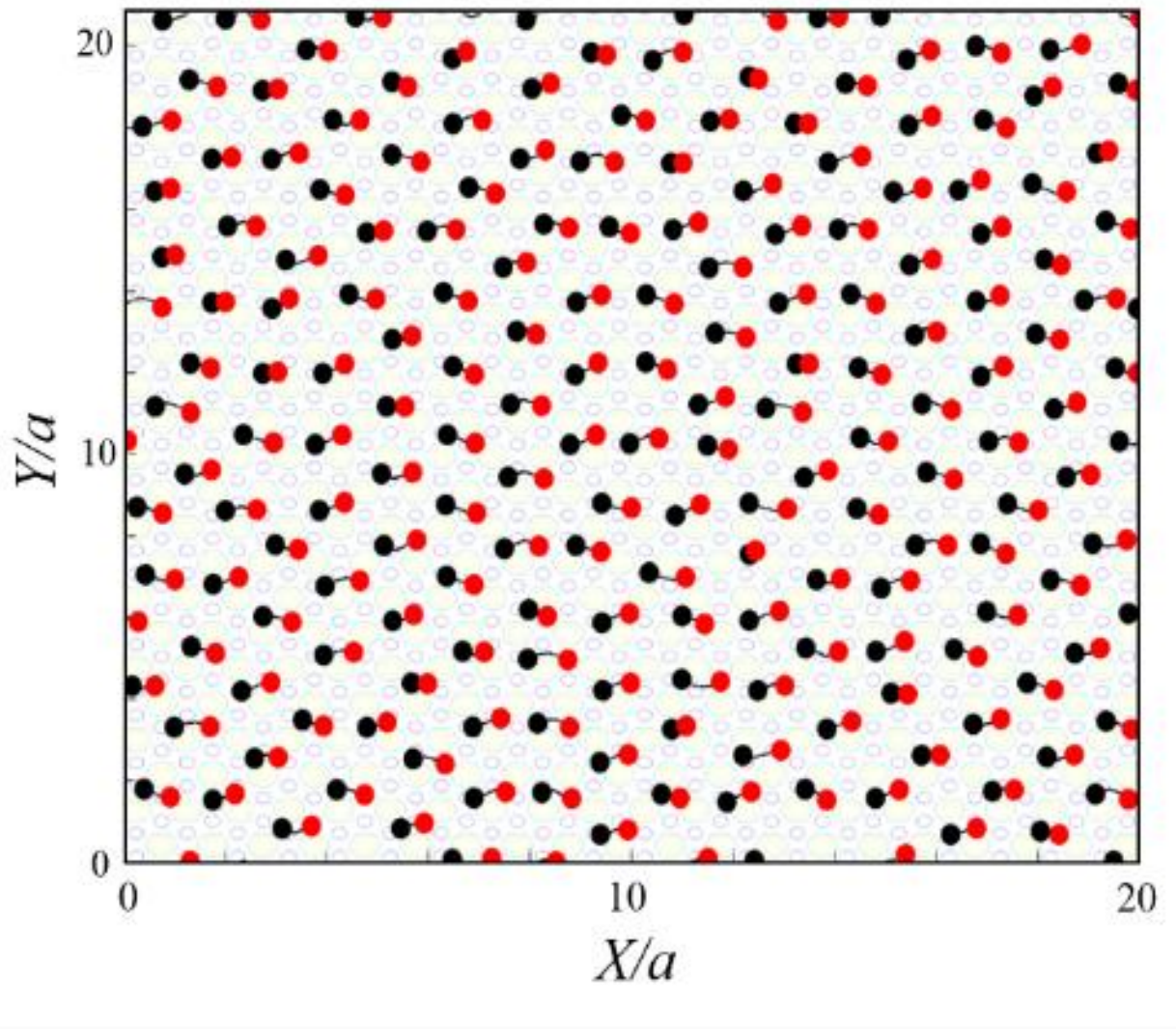

**Figure S6. 2D simulated Flux Flow.** Flux Flow is simulated using a value of $I=2.2$ and $f=0.20$. Unlike the defect motion regime shown in Figure 4 (b) with $I$=0.6, all vortices are in motion at once.

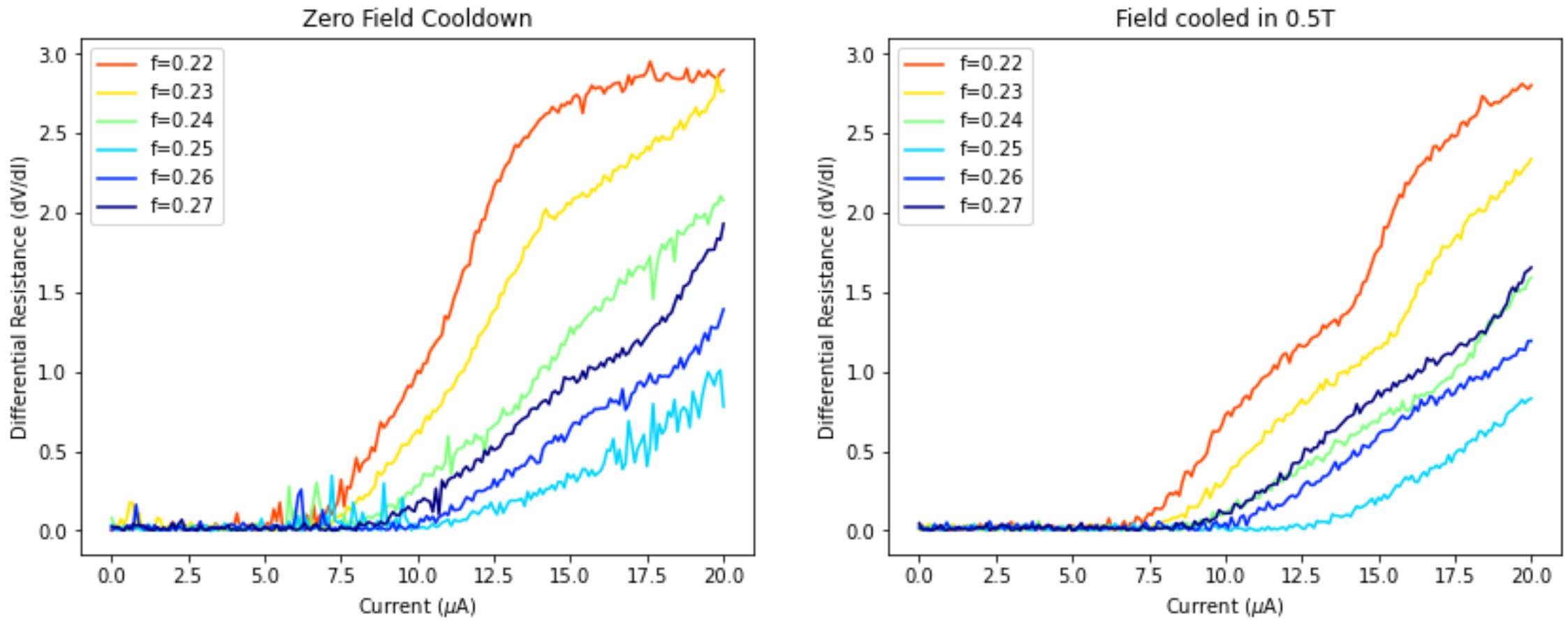

**Figure S7. Field cooled and zero field cooled dependence of superconducting island array. (a)** Differential resistance versus current at frustrations 0.22 through 0.27 on a superconducting island array with a lattice constant of 550 nm after cooling down the array with zero applied magnetic field. **(b)** Differential resistance versus current of the same island array as in (a), except the sample was cooled from 8K to 30 mK while a 0.5 T field was applied. Near commensurate filling ($f = 0.25$), we see little change in behavior. At more incommensurate fillings ($f = 0.22$, for example), there is a change in the shape due to shifts in the vortex lattice that occurred when some vortices were strongly pinned during field cooling.

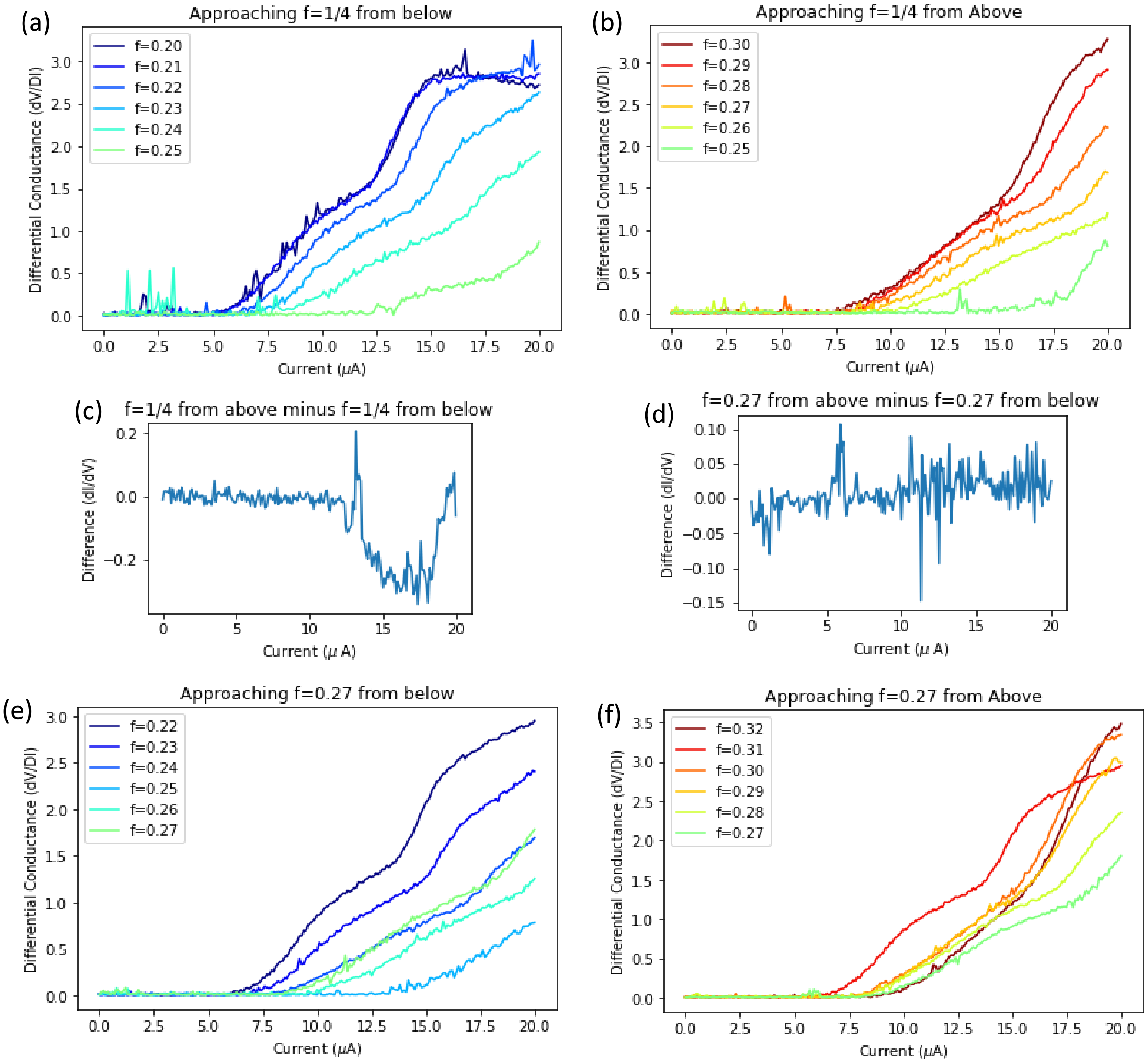

**Figure S8. History dependence on differential conductance (a)** Differential resistance taken starting at $f = 0.20$ and increasing the field to $f = 0.25$. **(b)** Differential resistance taken starting at $f = 0.30$ and decreasing the field to $f = 0.25$. **(c)** The difference in differential resistance of the $f = 0.25$ curve in (a) and (b). There is a shift in the depinning current, with the $f = 0.25$ curve as approached from higher fields having a larger depinning current. This is likely due to the vortex lattice being closer to exactly commensurate frustration due to remanent vortices from higher fields. **(d)** The difference in differential resistance of the $f = 0.27$ curve in (e) and (f). There appears to be no discernable difference when ramping the field up to an incommensurate frustration versus ramping the field down. **(e)** Differential resistance taken starting at $f = 0.22$ and increasing the field to $f = 0.27$. **(f)** Differential resistance taken starting at f=0.32 and decreasing the field to $f = 0.27$.